\documentclass[twocolumn,english,xlinenumbers,aps,twocolumn]{revtex4-1}
\usepackage[T1]{fontenc}
\usepackage[utf8]{inputenc}
\usepackage{color}
\usepackage{babel}
\usepackage{amstext}
\usepackage{graphicx}
\usepackage[unicode=true,pdfusetitle,
 bookmarks=true,bookmarksnumbered=false,bookmarksopen=false,
 breaklinks=false,pdfborder={0 0 0},backref=false,colorlinks=true]
 {hyperref}
\hypersetup{
 citecolor=blue}

\makeatletter
\usepackage{babel}

\makeatother

\begin{document}

\title{Dissipative Light Bullets in Passively Mode-Locked Semiconductor
Lasers}

\author{J. Javaloyes}

\affiliation{Departament de Fisica, Universitat de les Illes Baleares, C/Valldemossa
km 7.5, 07122 Mallorca, Spain}
\begin{abstract}
We demonstrate the existence of stable three dimensional dissipative
localized structures in the output of a laser coupled to a distant
saturable absorber. These phase invariant light bullets are individually
addressable and can be envisioned for three dimensional optical information
storage. An effective theory provides for an intuitive picture and
allows to relate their formation to the morphogenesis of static auto-solitons
and cellular patterns. The complexity incurred by the widely different
time scales present in the problem as well as by non-local couplings
that stem from the material degrees of freedom is circumvented by
the use of a multiple time-scale analysis. This provides a powerful
model enabling to tackle effectively the three dimensional case.
\end{abstract}
\maketitle
The possibility of using light bullets (LBs), i.e. pulses simultaneously
confined in the transverse and the propagation directions, has attracted
a lot of interest in the last twenty years, both for fundamental and
practical reasons. Optical confinement is ordinarily envisioned trough
a mechanism in which a Kerr self-focusing nonlinearity compensates
the spreading effect of diffraction. Yet in this scenario, reminiscent
of the conservative soliton theory, conservative LB are unstable and
lead to a collapse \cite{S-OL-90} in three dimensions. Other confinement
mechanisms were envisioned in systems far from equilibrium ande LBs
were predicted in semiconductor cavities \cite{VVK-OS-00,BMP-PRL-04}.
However, it was later shown that the proper consideration of the dynamics
of the active material leads to collapse \cite{CPM-NJP-06}.

In this letter, we establish the existence of stable LBs in semiconductor
laser cavities. Departing from previous approaches, the structuration
mechanism exploits the existence of the active material temporal scales
making our LB essentially multiple timescale objects. We present an
intuitive theory that allows to understand the LBs as hybrids between
the temporal localized structures described in \cite{MJB-PRL-14}
and the spatial diffractive autosolitons of \cite{RK-OS-88}.

Localized structures (LS) have been widely observed in nature \cite{WKR-PRL-84,MFS-PRA-87,NAD-PSS-92,UMS-NAT-96,AP-PLA-01}
and can be interpreted ---in the weak dissipative limit--- as dissipative
solitons \cite{TF-JPF-88,FT-PRL-90}. They may form when two homogeneous
solutions coexist for the same values of the parameters \cite{MMN-PRL-83,RK-OS-88,RK-JOSAB-90}.
These LS correspond to a locking between two opposed fronts connecting
the two stable states. Another case consists in the coexistence of
an homogeneous and a modulated solution \cite{TML-PRL-94} leading
to the so-called cellular patterns \cite{CRT-PRL-00}.

One separates in optics the systems in which the LS are locked to
an external injection beam from the ones that possess a phase invariance,
see \cite{L-CSF-94,MT-JOSAB-04} for a review. The former case leads
to the so-called cavity solitons \cite{FS-PRL-96,BLP-PRL-97} observed
either in the transverse plane of broad area amplifiers \cite{BTB-NAT-02}
or in the temporal output of fibers \cite{LCK-NAP-10,HBJ-NAP-14}.
In the latter case, diffractive autosolitons were predicted either
in situations where the cavity is composed of a gain medium coupled
to a saturable absorber \cite{RK-OS-88,RK-JOSAB-90} or to an external
diffraction grating \cite{TAF-PRL-08}. Because these \textit{lasing}
LS (LLS) appear in a phase invariant system \cite{GTB-EJPD-10}, their
properties are very different from phase-locked LS, leading for instance
to optical vortices \cite{GBG-PRL-10}.

It was recently shown \cite{MJB-PRL-14} that phase invariant temporal
LLS can evolve from passive mode-locking (PML). The PML regime leads
to the emission of temporal pulses much shorter than the cavity round-trip
\cite{haus00rev}. It is achieved by combining a laser amplifier providing
gain and a nonlinear loss element, usually a saturable absorber (SA).
The different dynamical properties of the two elements create a window
for regeneration only around the pulse. Although many pulses may coexist
within a PML cavity, they cannot be addressed individually. It is
due to the fact that the dynamics cannot be reduced to the sole evolution
of the field and while the pulses are typically short and may seem
temporally localized, they may still strongly coupled trough the evolution
of the gain that occurs on a much longer time scale.

For cavities with a large aspect ratio, as defined by the ratio of
the gain recovery time and of the cavity round-trip time $r=\tau/\tau_{g}$,
the PML pulses may become under certain conditions individually addressable
LLS coexisting with the off solution \cite{MJB-PRL-14}. In this regime
of temporal localization, the LLS were found to arise from a bifurcation
scenario radically different to the one found in PML lasers. While
PML occurs as a supercritical Andronov-Hopf bifurcation over a continuous
wave solution, here the LLS are nascent from Saddle Node bifurcations
of Limit cycles (SNL) and occur \emph{below} the lasing threshold.
As such, they coexist between themselves as well as with the zero
intensity solution. The coexistence with the harmonic PML solution
of maximal order hinted at a similarity with decomposable cellular
patterns \cite{CRT-PRL-00}. 

In this manuscript we discuss in which conditions the transverse profile
of these temporal LLS can self-organize yielding to a new robust LB
formation scenario. We describe the PML laser using the generic delayed
differential equation model presented in \cite{VT-PRA-05} which can
describe both the pulsating regimes and the steady solutions. We work
in the limit of low gain ($G$) and saturable absorption ($Q$) as
to justify a first order approximation to the single pass evolution
of the field profile. In addition, we assume that the transverse section
is sufficiently broad for the effect of diffraction at each round-trip
to be small. This allows the lumping and the commuting of the various
nonlinear elements. In this uniform field limit, the equation for
the field amplitude $E\left(r_{\perp},t\right)$ reads\begin{widetext}
\begin{eqnarray}
\left(\gamma^{-1}\partial_{t}+1-i\Delta_{\perp}\right)E\left(r_{\perp},t\right) & = & \sqrt{\kappa}\left[1+\frac{1-i\alpha}{2}G\left(r_{\perp},t-\tau\right)-\frac{1-i\beta}{2}Q\left(r_{\perp},t-\tau\right)\right]E\left(r_{\perp},t-\tau\right).\label{eq:VTJ1}
\end{eqnarray}

\end{widetext} where $\gamma$ is the bandwidth of the spectral filter,
$\Delta_{\perp}=\partial_{x}^{2}+\partial_{y}^{2}$ is the transverse
Laplacian, $\kappa$ is the fraction of the power remaining in the
cavity after each round-trip and $\alpha$ and $\beta$ are the linewidth
enhancement factors of the gain and absorber sections, respectively.
In Eq.~(\ref{eq:VTJ1}), the transverse space variables $r_{\perp}=\left(x,y\right)$
have been normalized to the diffraction length. As such, the domain
size $L_{\perp}$ representing the dimension of the broad area laser
becomes a bifurcation parameter: we foresee that with $L_{\perp}\ll1$,
one may only find a uniform spatial state while localized patterns
may occur when $L_{\perp}\gg1$. 

In addition to the two \emph{reversible} transverse spatial dimensions
$r_{\perp}$ (found setting $r_{\perp}\rightarrow-r_{\perp}$), the
delayed values of the variables $\left(E,G,Q\right)$ render Eq.~(\ref{eq:VTJ1})
infinite-dimensional along the propagation direction. The time delay
($\tau$) describes the spatial boundary conditions for a loop cavity
and governs the fundamental repetition rate of the PML laser. The
carrier equations read 
\begin{eqnarray}
\partial_{t}G & = & \Gamma G_{0}-G\left(\Gamma+\left|E\right|^{2}\right)+\mathcal{D}_{g}\Delta_{\perp}G,\label{eq:VTJ2}\\
\partial_{t}Q & = & Q_{0}-Q\left(1+s\left|E\right|^{2}\right)+\mathcal{D}_{q}\Delta_{\perp}Q,\label{eq:VTJ3}
\end{eqnarray}
with $G_{0}$ the pumping rate, $\Gamma=\tau_{g}^{-1}$ the gain recovery
rate, $Q_{0}$ is the value of the unsaturated losses which determines
the modulation depth of the SA and $s$ the ratio of the saturation
energy of the gain and of the SA sections while the scaled diffusion
coefficients are $\mathcal{D}_{g}$ and $\mathcal{D}_{q}$. In Eqs.~(\ref{eq:VTJ1}-\ref{eq:VTJ3})
time has been normalized to the SA recovery time. The slow temporal
scales associated with $G$ and $Q$ prevent their adiabatic elimination
which renders the dynamics along the propagation (time) axis irreversible,
notwithstanding the presence of dissipative terms.

The lasing threshold above which the off solution, defined as $\left(E,G,Q\right)=\left(0,G_{0},Q_{0}\right)$,
becomes unstable is $G_{th}=2/\sqrt{\kappa}-2+Q_{0}$. We study the
dynamical system defined in Eqs.~(\ref{eq:VTJ1}-\ref{eq:VTJ3})
in the newly found regimes of LLS \cite{MJB-PRL-14}, where the cavity
round-trip time is much longer than the gain recovery $r=\tau/\tau_{g}\gg1$
and for values of the gain below threshold, i.e. $G_{0}<G_{th}$.
In this regime, LLS occur essentially as multiple timescales objects
due to the widely different dynamics of the light and of the matter
components.

\begin{figure}[t]
\centering{}\includegraphics[width=1\columnwidth]{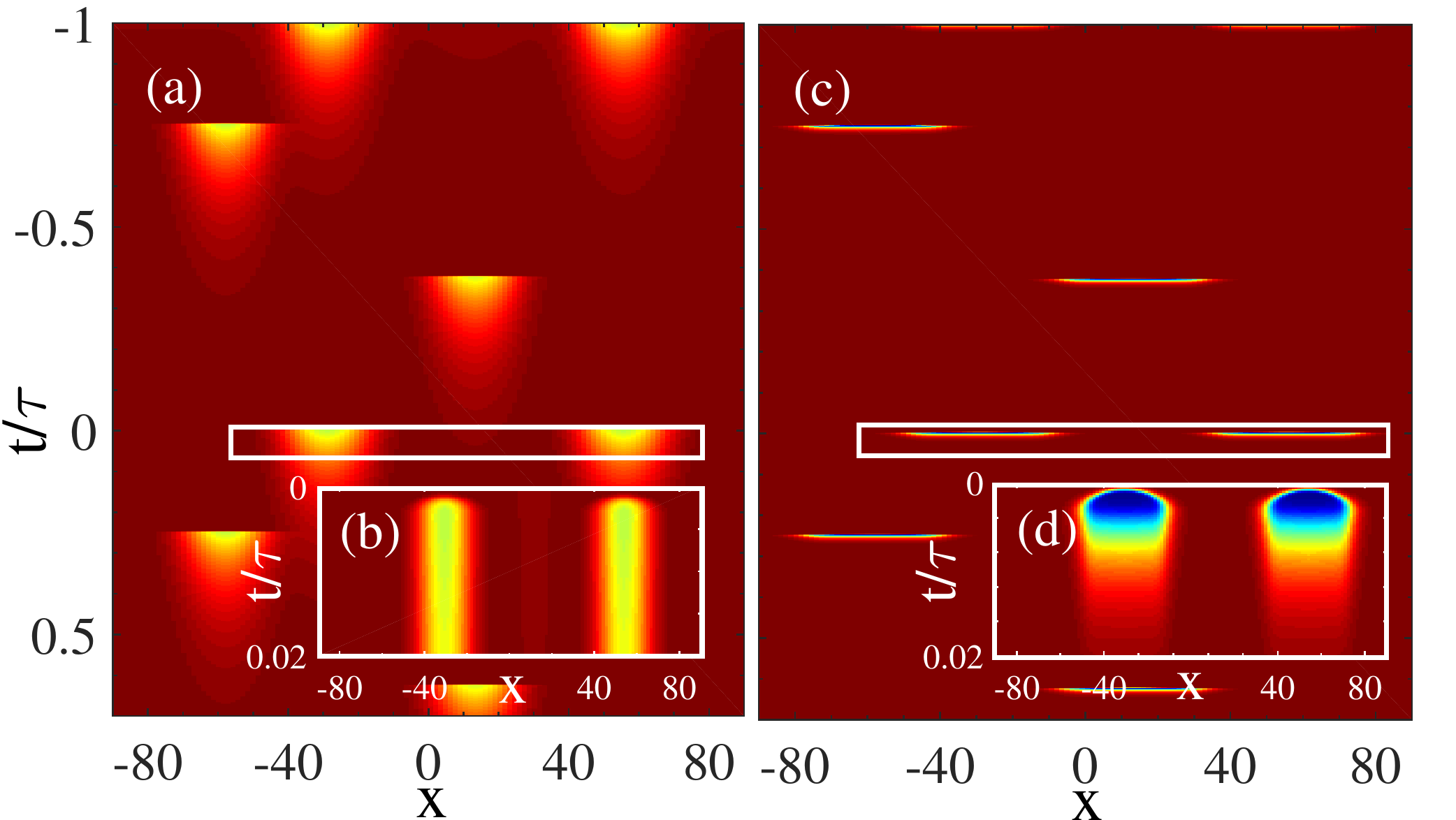} \caption{(Color online) Spatio-temporal trace for the gain $G\left(x,t\right)$
(a,c) and for the absorption $Q\left(x,t\right)$ (b,d) in a regime
where $N=3$ LBs are present. The time windows is such that each LB
appears twice. The inset (b,d) highlight the difference between the
relaxation timescales of the material variables. The slowest recovery
variables $G$ defines the temporal extent of the LB. The parameters
are $\gamma=40$, $\kappa=0.8$, $\alpha=\beta=0$, $\tau=200$, $G_{0}=0.6678G_{th}$,
$\Gamma=0.04$, $Q_{0}=0.3$, $s=30$, $L_{\perp}=90$ and $\mathcal{D}_{g,q}=0$.\label{fig1}
}
\end{figure}

\begin{figure}
\begin{centering}
\includegraphics[bb=0bp 0bp 470bp 280bp,clip,width=0.97\columnwidth]{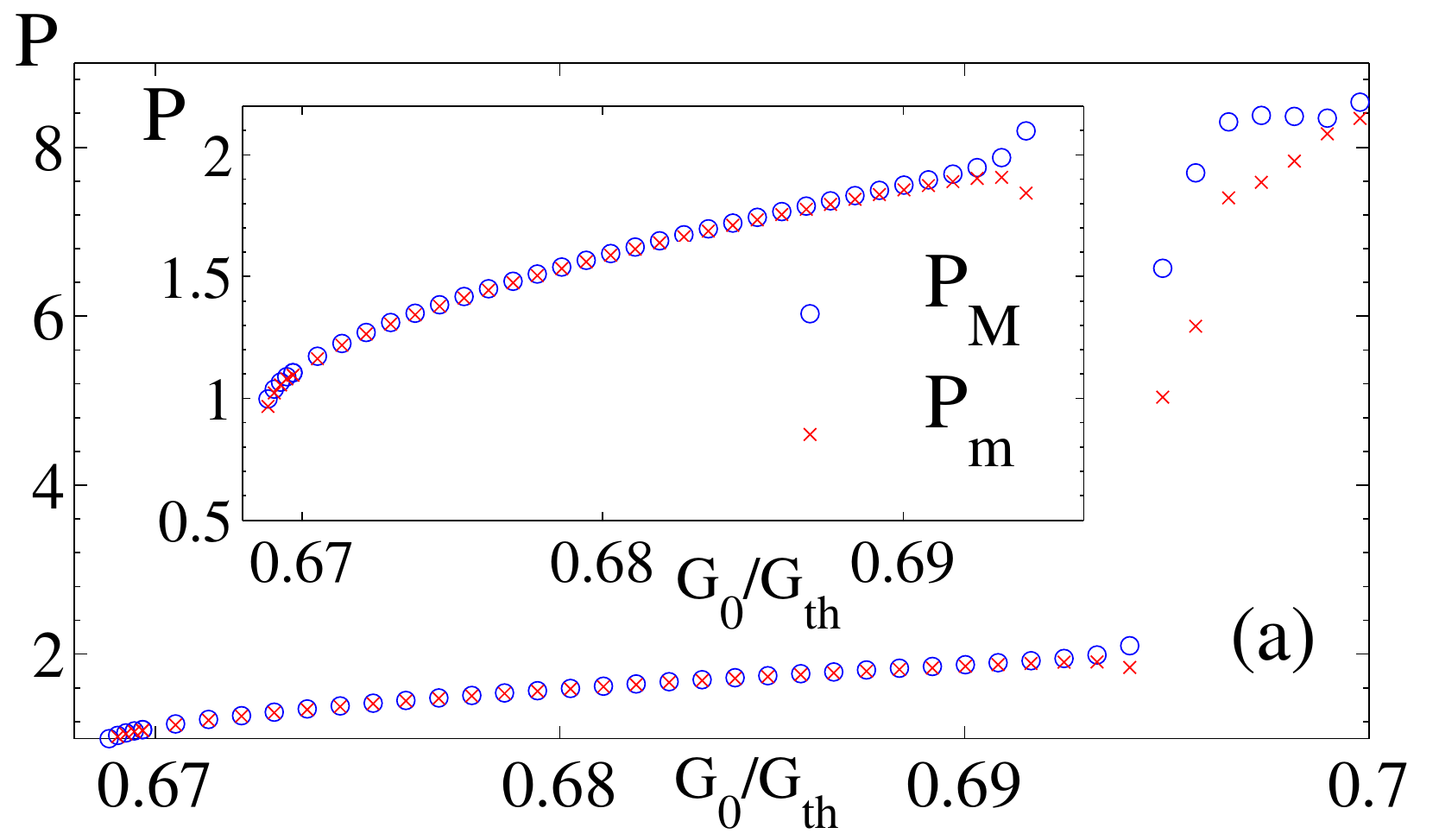}
\par\end{centering}

\centering{}\includegraphics[bb=0bp 0bp 480bp 280bp,clip,width=1\columnwidth]{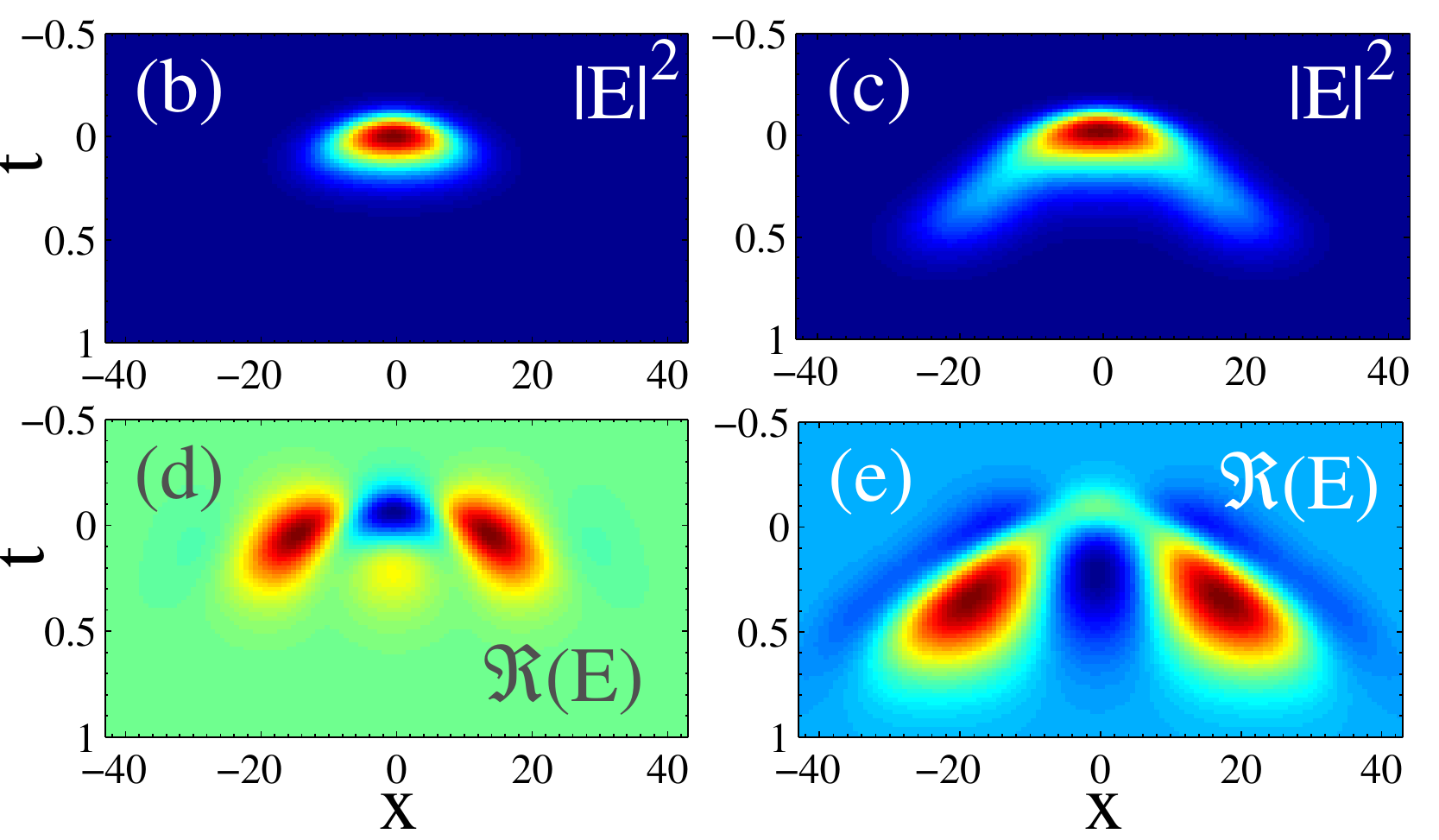}\caption{(Color online). Bifurcation diagram for the spatially integrated peak
power of the LB intensity ($P$) as a function of the gain $G_{0}$.
The single LB regime occurs as a SNL bifurcation around $G_{SN}=0.668G_{th}$
and looses its stability via an Andronov-Hopf bifurcation at $G_{H}\sim0.694G_{th}$.
In both cases, the spatio-temporal profile of the field is represented
in panels b,d) and c,e) with $G_{0}=G_{SN}$ and $G_{0}=G_{H}$, respectively.
The others parameters are like in Fig~\ref{fig1}.\label{fig2} }
\end{figure}

We show in Fig.~\ref{fig1} that Eqs.~(\ref{eq:VTJ1}-\ref{eq:VTJ3})
can indeed support the emission of LBs as temporal LLS that are also
confined in the transverse plane. Here the extent of the optical component
of the LB normalized to the round-trip time is typically $\tau_{p}/\tau\sim10^{-4}$.
Due to the very large aspect ratio of our system, such LBs would be
invisible in a spatio-temporal representation for the optical field
intensity as it would corresponds to an extremely thin horizontal
segment. It is why we represented in Fig.~\ref{fig1} the slower
material variables $G$ and $Q$. Starting from the off solution,
i.e. the solution with $N=0\,$LB, each additional LB was triggered
individually by sending a perturbation in the form of an optical pulse.
Since LBs are attractor, the details of the writing pulse are irrelevant
as long as its energy remains above a certain threshold. After a transient
of several tens of round-trips, the situation reaches an equilibrium
and an additional LB is written. With respect to the existing LBs,
a new LB can be written either at the same time but at a different
location, like the two LBs represented in the white inset in Fig.~\ref{fig1},
or a two different times. In this case, then can even partially overlap
spatially like the two leftmost LBs in Fig.~\ref{fig1} at coordinates
$\left(x,t/\tau\right)=\left(-40,-0.75\right)$. Without this staggering,
these two LBs could not coexist at the same instant.

In the case of a single LB, we performed a bifurcation diagram as
a function of the gain parameter. Our results are depicted in Fig.~\ref{fig2}
where we represent the pulse energy. Here one observes that the LBs
occurs below the lasing threshold. The square root behavior around
the minimal gain value suggests a scenario based upon a SNL bifurcation.
For high current, the LB spatial profile develops wings that oscillate
at a low frequency signaling the onset of an Andronov-Hopf bifurcation.
We depict in Fig.~\ref{fig1}b-e) the details of the LB at the lower
and highest stable currents. 

We stress that the self-organization mechanism at work is not based
upon a self-focusing Kerr nonlinearity. For instance, we obtained
the results of Fig.~\ref{fig1} and Fig.~\ref{fig2} setting the
values $\alpha=\beta=0$. Exploiting the seminal work of New \cite{N-JQE-74}
and the fact that the LLS are composed of different variables evolving
over widely different timescales, we find an effective equation for
the transverse morphogenesis problem in presence of diffraction. We
assume that the field reads $E\left(r_{\perp},t\right)=A\left(r_{\perp},t\right)p\left(t\right)$
with $p\left(t+\tau\right)=p\left(t\right)$ a short normalized temporal
pulse train and $A\left(r_{\text{\ensuremath{\perp}}},t\right)$ a
slowly evolving amplitude. Inserting this Ansatz in Eqs.~(\ref{eq:VTJ1}-\ref{eq:VTJ3}),
defining $\sigma=\gamma t$, we find that the equation governing the
dynamics of the transverse profile reads 
\begin{eqnarray}
\text{\ensuremath{\partial}}_{\sigma}A & = & i\Delta_{\perp}A+A\, f\left(\left|A\right|^{2}\right).\label{eq:Rosanov}
\end{eqnarray}
while the expression of the nonlinear function $f$ reads
\begin{eqnarray}
f\left(I\right) & = & \sqrt{\kappa}\left[1+\frac{1-i\alpha}{2I}G_{0}\left(1-e^{-I}\right)\right.\label{eq:f}\\
 &  & \left.-\frac{1-i\beta}{2sI}Q_{0}\left(1-e^{-sI}\right)\right]-1.\nonumber 
\end{eqnarray}

To obtain Eqs.~(\ref{eq:Rosanov},\ref{eq:f}), we used the fact
that, during the emission of a LLS, the stimulated terms are dominant
in Eqs.~(\ref{eq:VTJ2}-\ref{eq:VTJ3}) which allows connecting the
values of the material variables after the pulse impinging to the
values before. In our case where $\tau\gg\tau_{g}$, the material
variables perform a full recovery in-between the emission of a LLS.

\begin{figure}[b]
\centering{}\includegraphics[clip,width=1\columnwidth]{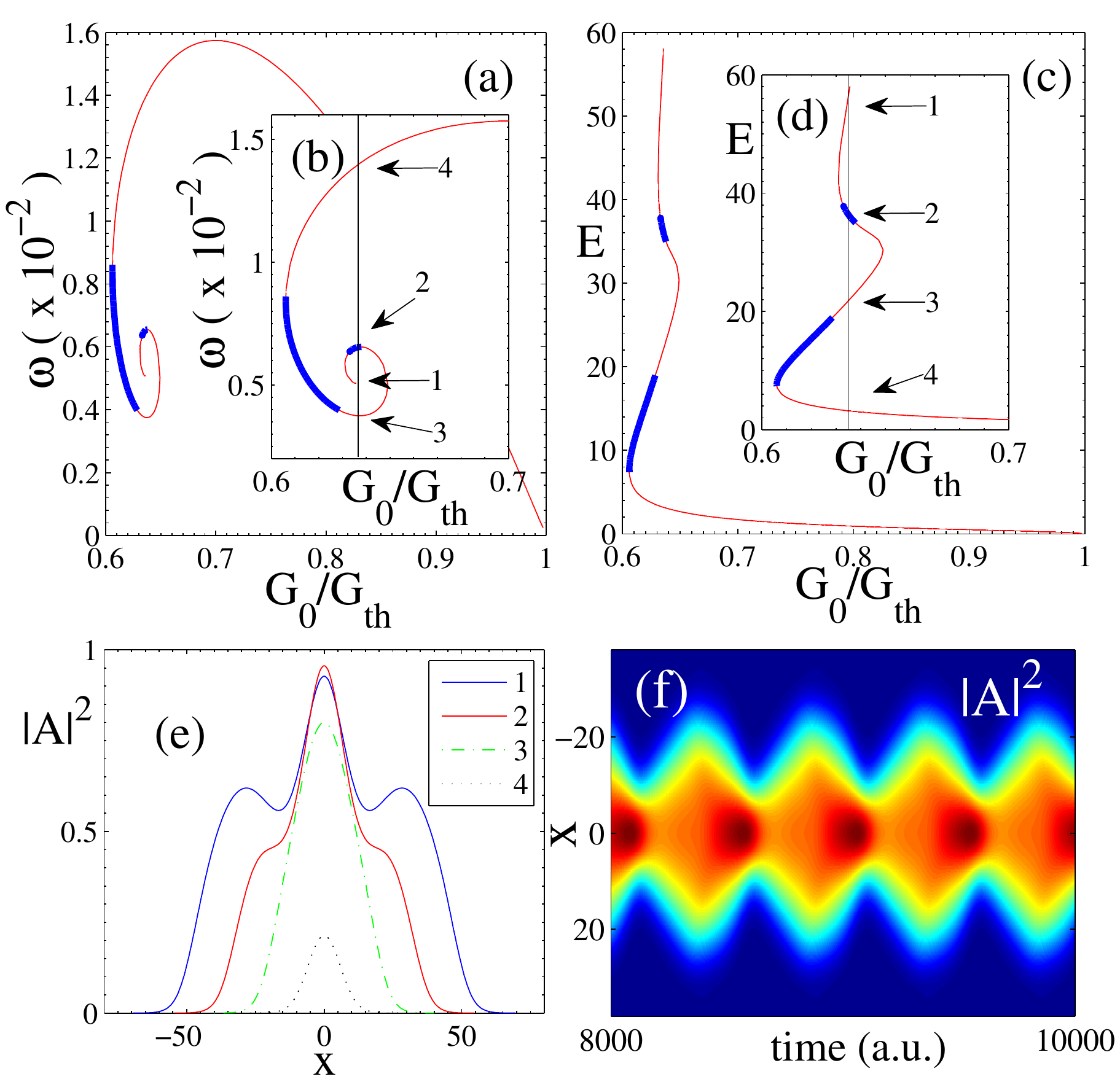}
\caption{Bifurcation diagram of the reduced model Eq.~(\ref{eq:Rosanov})
as a function of the gain parameter $G_{0}$. The panels a,b) represent
the evolution of the spectral parameter $\omega$ while the spatially
integrated intensity $\mathcal{E}$ is depicted in panels c,d). The
strong multistability of solutions is depicted in panel e). Usually,
only the transverse LS of higher intensity is stable. The main solution
is stable in the vicinity of the Saddle-Node bifurcation that occurs
at $G_{sn}=0.606G_{th}$ and losses its stability via and Andronov-Hopf
bifurcation at $G_{h}=0.63G_{th}$. We represent the oscillations
of the transverse profile slightly above $G_{0}=G_{h}$ in panel f).
Parameters as in Fig.~\ref{fig1}. \label{fig3} }
\end{figure}

We wrote Eq.~(\ref{eq:Rosanov}) to make apparent the link between
our approach and the work of \cite{rosanov,VFK-JOB-99} for the case
of static auto-solitons in bistable interferometers. In \cite{rosanov,VFK-JOB-99},
one assumes a monomode continuous wave emission along the longitudinal
propagation direction, which allows, via the adiabatic elimination
of the material variables, to find an effective equation for the transverse
profile, yet with a different expression for the function $f$. This
would corresponds in Eqs.~(\ref{eq:VTJ1}-\ref{eq:VTJ3}) to taking
the limit $\tau\rightarrow0$ and setting $\partial_{t}G=\text{\ensuremath{\partial}}_{t}Q=0$.
It is striking that in the frame of a temporal train of short pulsating
LLS, one finds exactly the same effective equation describing the
transverse structuration dynamics. Following the method detailed in
\cite{VFK-JOB-99} in the case of a single transverse spatial dimension,
we introduce in Eq.~(\ref{eq:Rosanov}) a phase-amplitude decomposition
and a spectral parameter $\omega$ as 
\begin{eqnarray}
A\left(x,\sigma\right) & = & \rho\left(x\right)\exp\left\{ i\left[\phi\left(x\right)-\omega\sigma\right]\right\} ,
\end{eqnarray}
and are left searching for static spatial LLS as heteroclinic and
homoclinic orbits of 
\begin{eqnarray}
\partial_{x}\rho & = & k\rho\;,\;\partial_{x}q=-2qk+\Re\left[f\left(\rho^{2}\right)\right],\\
\partial_{x}k & = & -\omega+q^{2}-k^{2}-\Im\left[f\left(\rho^{2}\right)\right],
\end{eqnarray}
where we defined $k=\rho^{-1}\partial_{x}\rho$ and $q=\partial_{x}\phi$.
Upon continuation as a function of a parameter one is able to reconstruct
a full bifurcation diagram and our results are shown in Fig.~\ref{fig3}.
Here, the spiral shape of the single LS branch is reminiscent of the
results of \cite{VFK-JOB-99} in the case of static auto-solitons.
The stability of the solution branch was found via the reconstruction
of the linear evolution operator along the lines of \cite{PJB-OE-11,GJT-NC-15}.
As detailed in \cite{VFK-JOB-99}, Eq.~(\ref{eq:Rosanov}) exhibits
three zero eigenvalues related to the phase, translational and Galilean
invariance. 
\begin{figure}
\centering{}\includegraphics[bb=0bp 0bp 480bp 280bp,clip,width=1\columnwidth]{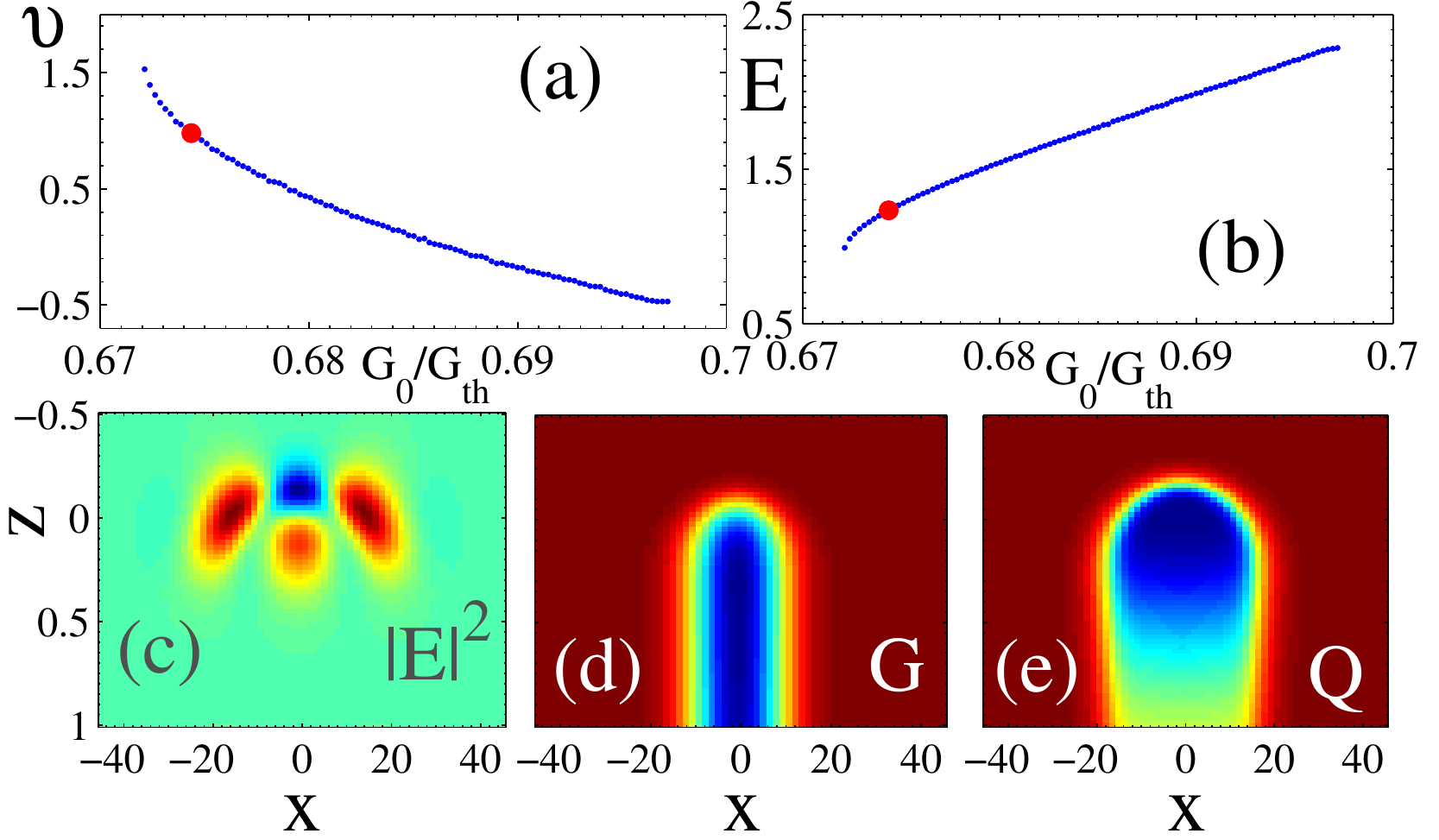}
\caption{Bifurcation diagram as a function of $G_{0}$. We represent the drift
velocity of the LB (a), their energy (b) as well as the field (c),
gain (d) and absorption (e) spatio-temporal profiles at $G_{0}=0.674G_{th}$,
indicated in panels a,b) by a circle.\label{fig4}}
\end{figure}
 This reduced model bifurcation diagram is in good qualitative agreement
with one of the full model. The extent of the stable LB region is
well reproduced as $\Delta G\sim0.03G_{th}$. The rigid shift to the
left in the reduced model can be ascribed to the neglect of the spectral
filtering loss term inherent to New's approach. Finally, the case
of radially symmetric 2D LBs can be studied similarly.

\begin{figure}[b]
\centering{} \includegraphics[bb=0bp 0bp 4325bp 1424bp,clip,width=1\columnwidth]{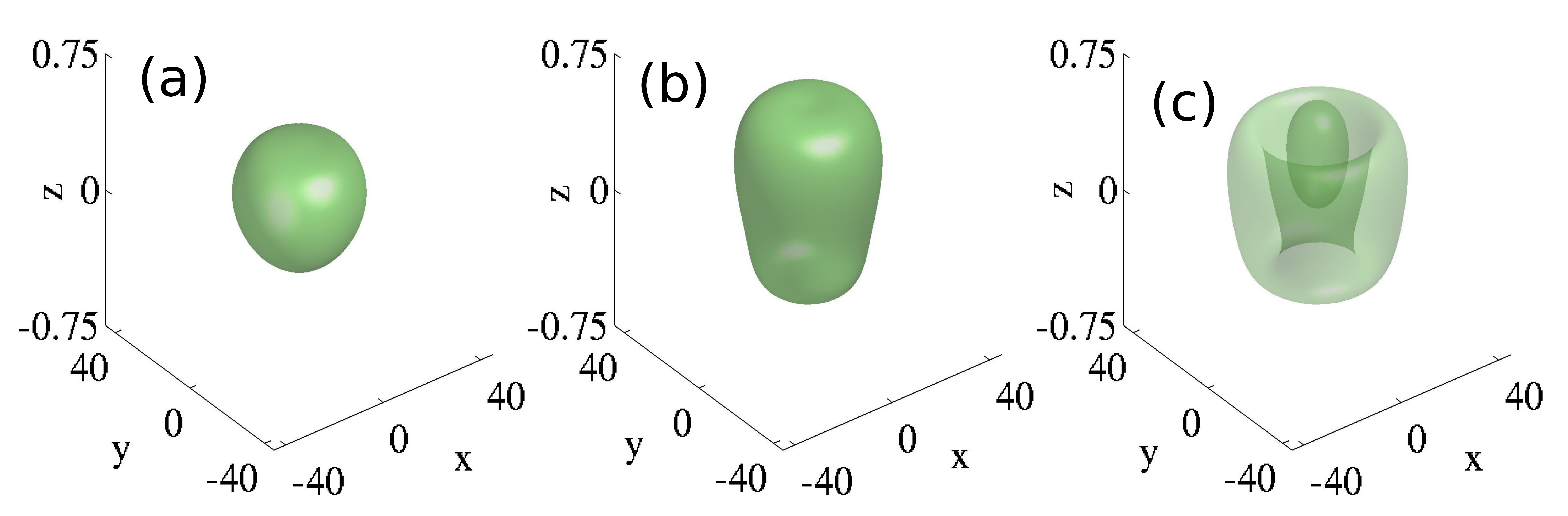}
\caption{Isosurface representation of a three dimensional LB at which the Intensity
(a), the real (b) and the imaginary parts (c) are $1\,$\% of the
maximal value. The multiple curves in (c) indicate that the LB are
spatially oscillating. \label{fig5} }
\end{figure}

However, the aforementioned reduced model does not allow access to
the temporal pulsewidth. One can transform Eqs.~(\ref{eq:VTJ1}-\ref{eq:VTJ3})
into a more tractable PDE using a multiple time scale analysis. In
the limit of a broadband filter $\gamma\rightarrow\infty$, we reduce
Eqs.~(\ref{eq:VTJ1}-\ref{eq:VTJ3}) to the Haus equation 
\begin{eqnarray}
\partial_{T}E & = & \left[\sqrt{\kappa}\left(1+\frac{1-i\alpha}{2}G-\frac{1-i\beta}{2}Q\right)-1\right]E,\nonumber \\
 & + & \left(\frac{1}{2\gamma^{2}}\partial_{z}^{2}+i\Delta_{\perp}\right)E\label{eq:VTJH-1}\\
\partial_{z}G & = & \Gamma G_{0}-G\left(\Gamma+\left|E\right|^{2}\right)+\mathcal{D}_{g}\Delta_{\perp}G,\label{eq:VTJH-2}\\
\partial_{z}Q & = & Q_{0}-Q\left(1+s\left|E\right|^{2}\right)+\mathcal{D}_{q}\Delta_{\perp}G,\label{eq:VTJH-3}
\end{eqnarray}
with $T$ a slow time scale. The Eqs.~(\ref{eq:VTJH-1}-\ref{eq:VTJH-3})
clarify the composite nature of the LBs as one notices the presence
of diffusion along the longitudinal dimension and diffraction in the
transverse dimensions. Although Eqs.~(\ref{eq:VTJH-1}-\ref{eq:VTJH-3})
may seems a complication with respect to Eqs.~(\ref{eq:VTJ1}-\ref{eq:VTJ3})
since the system is now 4D, it allows to cut the domain along the
$z$-axis to a box of the size of the \emph{optical pulse} and neglect
the long tail of the gain recovery. The reason for doing so is that
the gain material losses entirely its memory at the next round-trip,
so that one can always set the boundary condition $\left(G,Q\right)\left(z=0,r_{\perp},T\right)=\left(G_{0},Q_{0}\right)$.
One can appreciate in Fig.~\ref{fig4} the results of this approach
and notice the excellent agreement with the results of the full model.
Finally, this approach can readily be extended to the study of 3D
LB as depicted in Fig.~\ref{fig5}.

In conclusion, we have shown that broad area PML lasers with a large
temporal aspect-ratio can display addressable Lasing Light bullets.
We presented an intuitive theory that allows to understand these multiple
time scales object as hybrids between the temporal localized structures
\cite{MJB-PRL-14} and spatial diffractive auto-solitons \cite{RK-OS-88}.
A multiple time scale analysis allows reducing the size of the longitudinal
domain by several orders of magnitude rendering the three dimensional
analysis feasible.
\begin{acknowledgments}
I acknowledge support from the Ramón y Cajal program and project RANGER
(TEC2012-38864-C03-01).
\end{acknowledgments}


\begin{thebibliography}{10}

\bibitem{S-OL-90}
Y.~Silberberg.
\newblock Collapse of optical pulses.
\newblock {\em Opt. Lett.}, 15:1282--1284, 1990.

\bibitem{VVK-OS-00}
N.A. Veretenov, A.G. Vladimirov, N.A. Kaliteevskii, N.N. Rozanov, S.V. Fedorov,
  and A.N. Shatsev.
\newblock Conditions for the existence of laser bullets.
\newblock {\em Optics and Spectroscopy}, 89(3):380--383, 2000.

\bibitem{BMP-PRL-04}
M.~Brambilla, T.~Maggipinto, G.~Patera, and L.~Columbo.
\newblock Cavity light bullets: Three-dimensional localized structures in a
  nonlinear optical resonator.
\newblock {\em Phys. Rev. Lett.}, 93:203901, 2004.

\bibitem{CPM-NJP-06}
L.~Columbo, I.~M. Perrini, T.~Maggipinto, and M.~Brambilla.
\newblock 3d self-organized patterns in the field profile of a semiconductor
  resonator.
\newblock {\em New Journal of Physics}, 8(12):312.

\bibitem{MJB-PRL-14}
M.~Marconi, J.~Javaloyes, S.~Balle, and M.~Giudici.
\newblock How lasing localized structures evolve out of passive mode locking.
\newblock {\em Phys. Rev. Lett.}, 112:223901, Jun 2014.

\bibitem{RK-OS-88}
N.~N. Rosanov and G.~V. Khodova.
\newblock Autosolitons in nonlinear interferometers.
\newblock {\em Opt. Spectrosc.}, 65:449--450, 1988.

\bibitem{WKR-PRL-84}
J.~Wu, R.~Keolian, and I.~Rudnick.
\newblock Observation of a nonpropagating hydrodynamic soliton.
\newblock {\em Phys. Rev. Lett.}, 52:1421--1424, Apr 1984.

\bibitem{MFS-PRA-87}
E.~Moses, J.~Fineberg, and V.~Steinberg.
\newblock Multistability and confined traveling-wave patterns in a convecting
  binary mixture.
\newblock {\em Phys. Rev. A}, 35:2757--2760, Mar 1987.

\bibitem{NAD-PSS-92}
F.~J. Niedernostheide, M.~Arps, R.~Dohmen, H.~Willebrand, and H.~G. Purwins.
\newblock Spatial and spatio-temporal patterns in pnpn semiconductor devices.
\newblock {\em physica status solidi (b)}, 172(1):249 -- 266, 1992.

\bibitem{UMS-NAT-96}
P.~B. Umbanhowar, F.~Melo, and H.~L. Swinney.
\newblock Localized excitations in a vertically vibrated granular layer.
\newblock {\em Nature}, (382):793 -- 796, 1996.

\bibitem{AP-PLA-01}
Yuri~A. Astrov and H.G. Purwins.
\newblock Plasma spots in a gas discharge system: birth, scattering and
  formation of molecules.
\newblock {\em Physics Letters A}, 283(5–6):349 -- 354, 2001.

\bibitem{TF-JPF-88}
O.~Thual and S.~Fauve.
\newblock Localized structures generated by subcritical instabilities.
\newblock {\em J.\ Phys.\ France}, 49:1829--1833, 1988.

\bibitem{FT-PRL-90}
S.~Fauve and O.~Thual.
\newblock Solitary waves generated by subcritical instabilities in dissipative
  systems.
\newblock {\em Phys. Rev. Lett.}, 64:282--284, Jan 1990.

\bibitem{MMN-PRL-83}
D.~W. McLaughlin, J.~V. Moloney, and A.~C. Newell.
\newblock Solitary waves as fixed points of infinite-dimensional maps in an
  optical bistable ring cavity.
\newblock {\em Phys. Rev. Lett.}, 51:75--78, 1983.

\bibitem{RK-JOSAB-90}
N.~N. Rosanov and G.~V. Khodova.
\newblock Diffractive autosolitons in nonlinear interferometers.
\newblock {\em J. Opt. Soc. Am. B}, 7:1057--1065, 1990.

\bibitem{TML-PRL-94}
M.~Tlidi, P.~Mandel, and R.~Lefever.
\newblock Localized structures and localized patterns in optical bistability.
\newblock {\em Phys. Rev. Lett.}, 73:640--643, Aug 1994.

\bibitem{CRT-PRL-00}
P.~Coullet, C.~Riera, and C.~Tresser.
\newblock Stable static localized structures in one dimension.
\newblock {\em Phys. Rev. Lett.}, 84:3069--3072, Apr 2000.

\bibitem{L-CSF-94}
L.A. Lugiato.
\newblock Transverse nonlinear optics: Introduction and review.
\newblock {\em Chaos, Solitons and Fractals}, 4(8–9):1251 -- 1258, 1994.
\newblock Special Issue: Nonlinear Optical Structures, Patterns, Chaos.

\bibitem{MT-JOSAB-04}
P.~Mandel and M.~Tlidi.
\newblock Transverse dynamics in cavity nonlinear optics (2000–2003).
\newblock {\em Journal of Optics B: Quantum and Semiclassical Optics},
  6(9):R60, 2004.

\bibitem{FS-PRL-96}
W.~J. Firth and A.~J. Scroggie.
\newblock Optical bullet holes: Robust controllable localized states of a
  nonlinear cavity.
\newblock {\em Phys. Rev. Lett.}, 76:1623--1626, Mar 1996.

\bibitem{BLP-PRL-97}
M.~Brambilla, L.~A. Lugiato, F.~Prati, L.~Spinelli, and W.~J. Firth.
\newblock Spatial soliton pixels in semiconductor devices.
\newblock {\em Phys. Rev. Lett.}, 79:2042--2045, 1997.

\bibitem{BTB-NAT-02}
S.~Barland, J.~R. Tredicce, M.~Brambilla, L.~A. Lugiato, S.~Balle, M.~Giudici,
  T.~Maggipinto, L.~Spinelli, G.~Tissoni, T.~Knodl, M.~Miller, and R.~Jager.
\newblock Cavity solitons as pixels in semiconductor microcavities.
\newblock {\em Nature}, 419(6908):699--702, Oct 2002.

\bibitem{LCK-NAP-10}
F.~Leo, S.~Coen, P.~Kockaert, S.P. Gorza, P.~Emplit, and M.~Haelterman.
\newblock Temporal cavity solitons in one-dimensional kerr media as bits in an
  all-optical buffer.
\newblock {\em Nat Photon}, 4(7):471--476, Jul 2010.

\bibitem{HBJ-NAP-14}
T.~Herr, V.~Brasch, J.~D. Jost, C.~Y. Wang, N.~M. Kondratiev, M.~L. Gorodetsky,
  and T.~J. Kippenberg.
\newblock Temporal solitons in optical microresonators.
\newblock {\em Nature Photonics}, 8(2):145--152, 2014.

\bibitem{TAF-PRL-08}
Y.~Tanguy, T.~Ackemann, W.~J. Firth, and R.~J\"ager.
\newblock Realization of a semiconductor-based cavity soliton laser.
\newblock {\em Phys. Rev. Lett.}, 100:013907, Jan 2008.

\bibitem{GTB-EJPD-10}
P.~Genevet, M.~Turconi, S.~Barland, M.~Giudici, and J.~R. Tredicce.
\newblock Mutual coherence of laser solitons in coupled semiconductor
  resonators.
\newblock {\em Eur. Phys. J. D}, 59:109--114, 2010.

\bibitem{GBG-PRL-10}
P.~Genevet, S.~Barland, M.~Giudici, and J.~R. Tredicce.
\newblock Bistable and addressable localized vortices in semiconductor lasers.
\newblock {\em Phys. Rev. Lett.}, 104:223902, Jun 2010.

\bibitem{haus00rev}
H.~A. Haus.
\newblock Mode-locking of lasers.
\newblock {\em IEEE J. Selected Topics Quantum Electron.}, 6:1173--1185, 2000.

\bibitem{VT-PRA-05}
A.~G. Vladimirov and D.~Turaev.
\newblock Model for passive mode locking in semiconductor lasers.
\newblock {\em Phys. Rev. A}, 72:033808, Sep 2005.

\bibitem{N-JQE-74}
G.~New.
\newblock Pulse evolution in mode-locked quasi-continuous lasers.
\newblock {\em Quantum Electronics, IEEE Journal of}, 10(2):115 -- 124,
  February 1974.

\bibitem{rosanov}
N.~N. Rosanov and G.~V. Khodova.
\newblock Autosolitons in bistable interferometers.
\newblock {\em Opt. Spectrosc.}, 65:449, 1988.

\bibitem{VFK-JOB-99}
A.~G. Vladimirov, S.~V. Fedorov, N.~A. Kaliteevskii, G.~V. Khodova, and N.~N.
  Rosanov.
\newblock Numerical investigation of laser localized structures.
\newblock {\em Journal of Optics B: Quantum and Semiclassical Optics},
  1(1):101, 1999.

\bibitem{PJB-OE-11}
A.~P\'{e}rez-Serrano, J.~Javaloyes, and S.~Balle.
\newblock Longitudinal mode multistability in {R}ing and {F}abry-{P}\'{e}rot
  lasers: the effect of spatial hole burning.
\newblock {\em Opt. Express}, 19(4):3284--3289, Feb 2011.

\bibitem{GJT-NC-15}
B.~Garbin, J.~Javaloyes, G.~Tissoni, and S.~Barland.
\newblock Topological solitons as addressable phase bits in a driven laser.
\newblock {\em Nat. Com.}, 6, 2015.

\end{thebibliography}

\end{document}